\documentclass[%
 aip,
 apl,%
 amsmath,amssymb,
preprint,%
 numerical%
]{revtex4-1}

\usepackage{graphics,graphicx}
\usepackage{dcolumn}
\usepackage{bm}

\begin{document}

\preprint{AIP/123-QED}

\title[Diamond Elasticity at HPT]{Elasticity of Diamond at  High Pressures and Temperatures
}

\author{Maribel N\'u\~nez Valdez}
\email{nunez025@umn.edu}
\affiliation{%
Dept. of Chemical Engineering and Materials Science, University of Minnesota, Minneapolis, Minnesota 55455, USA
}
\author{Koichiro Umemoto}%
\affiliation{ 
Dept. of Geology and Geophysics, University of Minnesota, Minneapolis, Minnesota 55455, USA
}%
\author{Renata M. Wentzcovitch}
\affiliation{%
Dept. of Chemical Engineering and Materials Science, University of Minnesota, Minneapolis, Minnesota 55455, USA
}
\affiliation{%
Minnesota Supercomputing Institute (MSI), University of Minnesota, Minneapolis, Minnesota 55455, USA
}
\date{\today}

\begin{abstract}
We combine density functional theory within the local density approximation, the quasiharmonic approximation, and  vibrational density of states to calculate single crystal elastic constants, and bulk and shear moduli of diamond  at simultaneous high pressures and temperatures in the ranges of $0-500$ GPa and $0-4800$ K. Comparison with experimental values at ambient pressure and high temperature shows an excellent agreement for the first time with our first-principles results validating our method. We show that the anisotropy factor of diamond increases to 40\% at high pressures and becomes temperature independent.
\end{abstract}

\maketitle

Diamond has singular physical properties. Its strength and transparency make it remarkably useful for industrial applications in abrasive machining, micro electromechanical devices etc. \cite{Kohn99} It is also an anvil material for experimental studies under extreme pressures and temperatures.  Pressures greater to that reached at the center of the Earth ($\sim364$ GPa) can be produced in diamond anvil cells (DACs) using single crystal (SC) diamonds as opposing anvils. Applications of diamond to geoscience and planetary science are ideal to study high pressure-temperature behavior of magmas and minerals relevant to the Earth and other planets. For example, in 2004 a high pressure form of (Fe,Mg)SiO$_3$, postperovskite, was discovered using DAC techniques \cite{Mura04}, this mineral exists above the mantle-core boundary of the Earth.

Natural diamond can be SC or polycrystalline (PC). The high quality SC diamond is optically transparent and often used as gems. It also possesses higher strength than the PC type, which is more opaque or translucent with impurities and inclusions. PC diamond, however, is ideal for cutting tools due to its toughness. 

Thin films of PC diamond with very low growth rate ($\sim1\mu$m/h) were synthesized and reported in the 1960s using chemical vapor deposition (CVD), but those films were of poor quality. SC diamond synthesis is performed near graphite-diamond phase boundary conditions ($\sim5-6$ GPa and $>1500$ K) within large-volume presses \cite{Shingley02}. The main limitations to this production are the size of the press, yielding only diamonds of some millimeters and a very slow growth rate \cite{Sumi02}. However, in recent years SCs of diamonds larger than 2 cm and more than 10 carats have been successfully fabricated using microwave plasma CVD techniques. The purest CVD diamond was reported to have comparable features to those of high purity natural SC with hardness and toughness that can be tuned by controlling growth and annealing \cite{Liang09,Meng12}. In 2003 in Japan, using large-volume presses \cite{Irifu03}, graphite was directly converted into diamond to form nanopolycrystalline diamond (NPD) at simultaneous high pressure and temperature ($\sim15$ GPa and $>2600$ K) within minutes. NPD was found to have high elastic stiffness, very high fracture toughness and similar or higher hardness than that of most single crystal diamonds. These features and the fact that NPD rods can be fabricated up to 1 cm in diameter and length (14.5 carats), make NPD very suitable for  larger anvils than the usually available in DACs. Larger sample volume in high pressure apparatuses has open new windows into mineral physics studies of materials that require accurate and simultaneous very high pressures and temperatures such as those found deep in the Earth and other planets \cite{Ohfuji10}. 

With these new techniques to fabricate diamonds and their increasing applications,  accurate elasticity of diamond at simultaneous high pressures and temperatures acquires new significance. However, it cannot be found in the literature in neither experimental form nor computational modeling. Using Brillouin scattering, elastic moduli of diamond were measured as a function of temperature in the range of 300--1600 K at ambient pressure \cite{Zoub98}. A molecular dynamics study attempted the computation of elastic constants between 100 and 1100 K with a modest degree of success \cite{Gao06}. Single crystal elasticity and bulk modulus of diamond were computed as a function of pressure at 300 K in the range of 0--500 GPa \cite{Fu09}, using first-principles methods based on density functional theory (DFT) \cite {HK,KS} as implemented in the CRYSTAL03 (CR03) program \cite{Saun03} and a qusiharmonic Debye model (QHDM) \cite{Blanco04}. But the  lack of experimental and other computational values for such quantities made impossible a comparison for validation of these CRYSTAL03 results.

In this letter, for the first time we address the elasticity of single crystal diamond and of a polycrystalline aggregate at simultaneous high pressures and temperatures. The computational approach is based on DFT \cite {HK,KS} within the local density approximation (LDA) \cite{Cep1980}. At arbitrary pressures, cubic equilibrium structures  of diamond (space group Fd3m, two C atoms/primitive cell)  were found using the variable cell-shape damped molecular dynamics approach  \cite{Wentz1991, Wentz1993} as implemented in the quantum-ESPRESSO (QE) code \cite{Giann}. The C pseudopotential was of Vanderbilt-type with cutoff radii of 1.3 a.u., and partial core correction. We used a plane-wave kinetic energy cutoff of 50 Ry and 200 Ry for the charge density. The {\bf k}-point sampling for charge density was determined on a $4\times4\times4$ Monkhorst-Pack grid of the Brillouin Zone (BZ) shifted by $\left(1/2,1/2,1/2\right)$ from the origin. These parameters corresponded to having interatomic forces and stresses smaller than $10^{-4}$ Ry/a.u. 

To obtain the independent cubic static elastic constants $C_{11}$, $C_{12}$, and $C_{44}$ (in Voigt notation \cite{Wallace}) at each pressure, the equilibrium structures were strained, and their internal degrees of freedom re-relaxed. Then elastic constants were extracted using the relationship 
$\sigma_{ij}=C_{ijkl}\varepsilon_{kl}$.
Positive and negative strains of 1\% magnitude were applied in order to attain accuracy in the limit of zero strain. This method has been applied successfully for over a decade \cite{Wentz1995,Karki1997,Nunez10,Nunez11}.

Dynamical matrices were calculated using density functional perturbation theory (DFPT) \cite{Baroni2001}. At each pressure, a dynamical matrix was obtained on a $4\times4\times4$ {\bf q}-point mesh. Force constants were extracted and interpolated on a $12\times12\times12$ mesh to produce VDoS. 

\begin{figure}
\includegraphics[width=80mm,height=157mm]{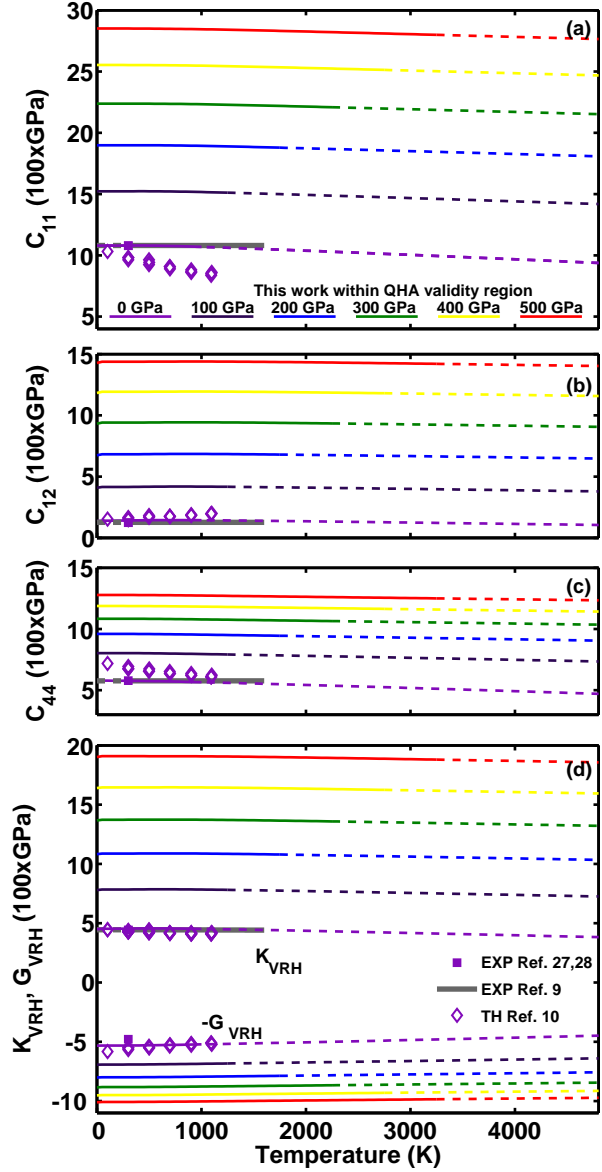}
\caption{\label{fig_1} (Color online) Temperature dependence of (a) $C_{11}$; (b) $C_{12}$; (c) $C_{44}$; and (d) bulk, $K$, and shear, $G$, moduli. Dashed lines indicate a lesser degree of confidence within the QHA limit.}
\end{figure}
\begin{table*}
\caption{\label{table1} Comparison of elastic mouduli of diamond at ambient conditions ($P$=0 GPa and $T$= 300K)} 
\begin{tabular}{ l l l l l l }
\hline
\hline
  Reference & $C_{11}$ (GPa) & $C_{12}$ (GPa)& $C_{44}$ (GPa)& $K_{VRH}$  (GPa)& $G_{VRH}$ (GPa)\\
\hline
 This work LDA & 1078.99 & 140.81 & 577.37 & 453.54 & 531.31 \\ 
   GGA \cite{Mounet05} &1060 & 125&562 & 436.67 & 522.08 \\
MD  \cite{Gao06} & 963.6-982.2 & 148.4-165.8 & 674.5-694.0 & 425.76-437.93 & 550.12-565.99 \\
 B3PW \cite{Fu09} & 1097.5 & 115.5 & 598.2 & 442.8 & 552.74 \\
  Exp \cite{McSk72}& 1079 & 124 & 578 & 442 & 535 \\
 Exp \cite{Grims75} & 1076.4 $\pm$ 0.2 & 125.2 $\pm$ 2.3 & 577.4 $\pm$ 1.4& 442.3 & 534.27\\
 Exp \cite{Zoub98} & 1080.4 & 127.0 & 576.6 & 444.8 & 534.32\\
 \hline
\hline
 \end{tabular}
\end{table*}
Then we exploited the dependence of phonon frequencies on compression to determine the thermal contribution to the Helmholtz free energy $F$, within the QHA \cite{Wallace}, 
\begin{eqnarray}
F\left(e,V,T\right)=U_{st}(e,V)+\frac{1}{2}\sum_{{\bf q},m}{\hbar\omega_{{\bf q},m}(e,V)}+\nonumber\\
+k_BT\sum_{{\bf q},m}\ln\left\{1-exp\left[-\frac{\hbar\omega_{{\bf q},m}(e,V)}{k_BT}\right]\right\},
\end{eqnarray}
where ${\bf q}$ is the phonon wave vector, $m$ is the normal mode index,  $T$ is temperature, $U_{st}$ is the static internal energy at equilibrium volume $V$ under isotropic pressure $P$ and infinitesimal strain $e$, $\hbar$ and $k_B$ are Planck and Boltzmann constants, respectively. Isothermal elastic constants are given by $C_{ijkl}^T=\left[\partial^2G(P,T)/\partial e_{ij}\partial e_{kl}\right]_{P,T}$ with $G=F+PV$, the Gibbs energy and $i,j,k,l=1,\dots,3$. To convert to adiabatic elastic constants, we use $C_{ijkl}^S=C_{ijkl}^T+\left(T/VC_V\right)\left(\partial S/\partial e_{ij}\right)\left(\partial S/\partial e_{kl}\right)\delta_{ij}\delta_{kl}$, ($C_V=$heat capacity at constant $V$, and $S=$entropy).  $C_{ijkl}^T$ have the static-$C_{ijkl}$ and phonon-$C_{ijkl}$ contributions, the latter can be expressed as function of strain and mode Gr\"unesein parameters. Assuming that the angular distribution of phonons is isotropic, isothermal elastic constants can be calculated without performing phonon calculations for strained configurations \cite{Wu11}. This approximation is equivalent to assuming that thermal pressure is isotropic. Even though this assumption is not completely rigorous\cite{Carrier07}, the method is quite accurate compared to experimental uncertainties \cite{Carrier08}. Bulk and shear moduli were calculated as Voigt-Reuss-Hill averages \cite{Watt} using the adiabatic $C_{ij}$, i.e., $K_{VRH}=(K_V+K_R)/2$ and $G_{VRH}=(G_V+G_R)/2$, where$K_V=K_R=(C_{11}+2C_{12})/3$, $G_V=(C_{11}-C_{12}+C_{44})/5$, and $G_R=15\left[12/(C_{11}-C_{12})+9/C_{44})\right]^{-1}$.  

The computed elastic constants $C_{11}$, $C_{12}$, and $C_{44}$, and bulk $K_{VRH}$ and shear $G_{VRH}$, moduli as a function of temperature are shown in Figure \ref{fig_1}. We compared our results with experimental trends in the temperature range of $300-1700$ K at ambient pressure and we found excellent agreement. The previous molecular dynamics calculation \cite{Gao06} was quite unsuccessful at reproducing the experimental values for $C_{ij}$ individually. The quality of our findings with respect to measurements validates our overall computational approach. 

\begin{figure}
\includegraphics[width=80mm,height=106mm]{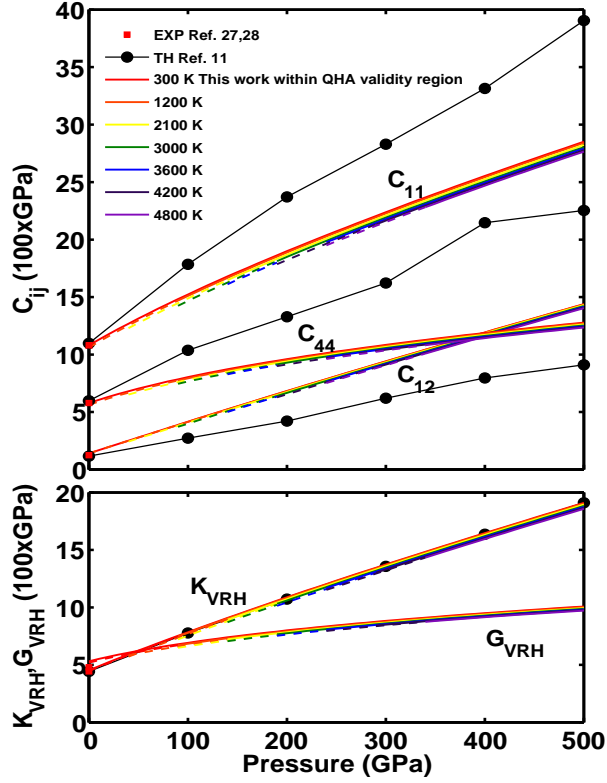}
\caption{\label{fig_2} (Color online) Pressure dependence of (a) $C_{11}$, $C_{12}$, and $C_{44}$; and (b) bulk, $K_{VRH}$, and shear, $G_{VRH}$, moduli. Dashed lines indicate a lesser degree of confidence within the QHA limit.}
\end{figure}

Then we obtained the elasticity as a function of pressure between 0 and 500 GPa at several temperatures, (Figure \ref{fig_2}). Unfortunately experimental values are only available at ambient conditions for comparison. The only other calculation of elasticity as a function of pressure \cite{Fu09} reported using CR03 and the QHDM was compared only to measurements at room temperature. Figure \ref{fig_2}(a) shows that CR03 results are quite different from ours. There is  agreement in the bulk modulus trend  [Figure \ref{fig_2}(b)], but that probably reflects the fact that the compression curves are similar in both studies. 

\begin{figure}
\includegraphics[width=75mm,height=148mm]{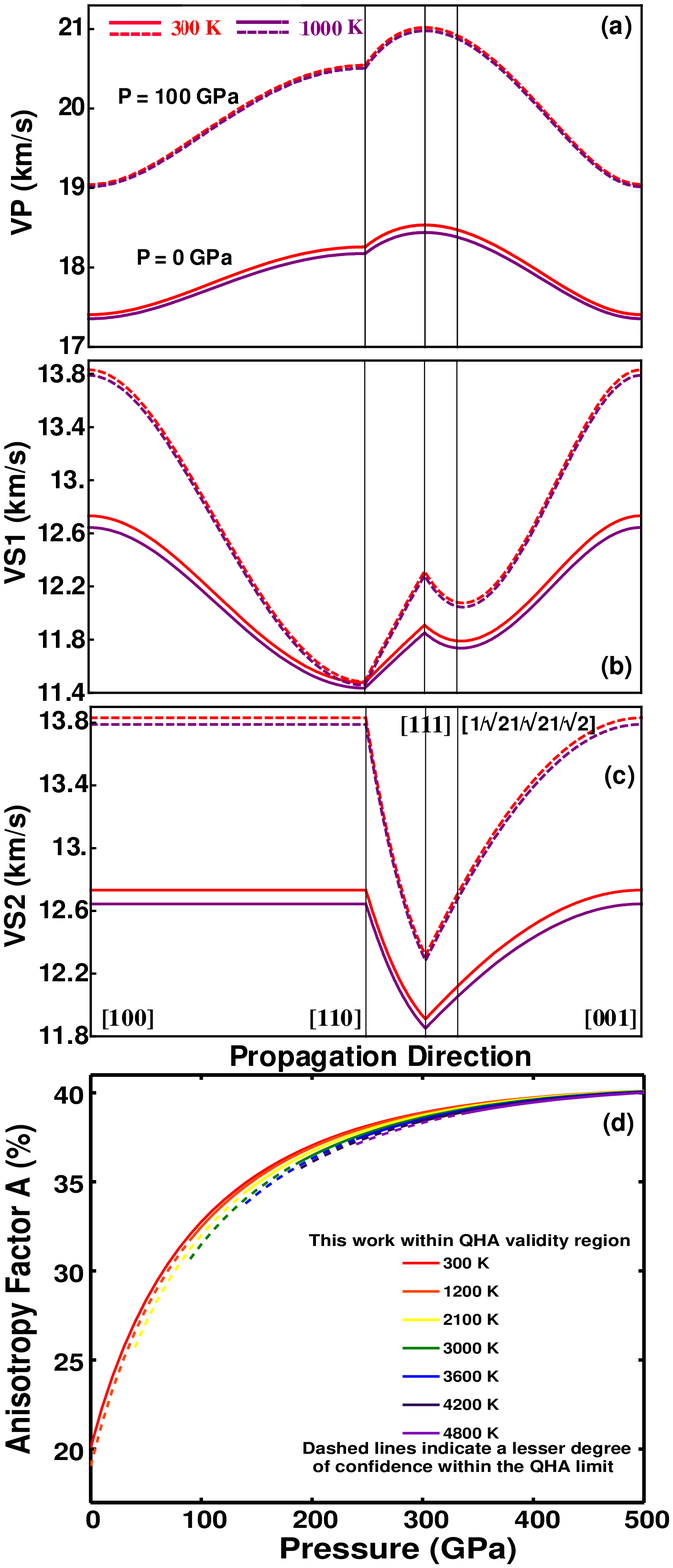}
\caption{\label{fig_3} (Color online) Variation of (a) Compressional velocity and (b,c) Shear velocities as a function of direction; and (d) Anisotropy factor of diamond}
\end{figure}

Since the QHA is valid up to about two thirds of the melting temperature (for diamond $T_{melt}=3823$ K at $P=0$ GPa and it increases in the range $0-500$ GPa \cite{Correa06}), dashed-lines in Figures \ref{fig_1} and \ref{fig_2} indicate a lesser degree of confidence for the QHA results \cite{Wu10}. Table \ref{table1} compares results on the elasticity of diamond at ambient conditions with the other DFT-based values and experimental measurements. One can see clearly the improvement in the computation of $C_{ij}$, $K_{VRH}$, and $G_{VRH}$ with respect to the other computational methods. Our computed bulk modulus is 2.5\% larger than experimental values. This is typical of LDA results that underestimate volume by $\sim1$\% and reflects on $C_{12}$ only. 
Finally, we solved the Christoffel equation \citep{Landau} for acoustic velocities in single crystal,  $\det|C_{ijkl}n_jn_l-\rho V^2\delta_{ik}|=0$, where $V$ is elastic wave velocity, $\mathbf{n}$ is propagation direction, $\delta_{ij}$ is Kronecker delta, and $\rho$ is density.
For a given $\mathbf{n}$, there are three solutions, i.e., one P-wave ($V_{P}$) and two S-waves ($V_{S1}$, $V_{S2}$). Figures \ref{fig_3}(a-c) show the variation with crystallographic direction (anisotropy) of $V_P$, $V_{S1}$ and $V_{S2}$ at 0 and 100 GPa and 300 and 1000 K for comparison. We found  similar qualitative behavior independent of pressure and temperature for the three waves. In Figure \ref{fig_3}(d) we show the anisotropic factor of diamond $A\left[(2C_{44}+C_{12})/C_{11}-1\right]\times100$, as a function of $P$ at several $T$s. One can observe that $A$ becomes roughly independent of $T$ and $P$.  $A$ often decreases with $P$ \cite{Karki99}, but this is not the case for diamond. As seen in Figures \ref{fig_3}(a-c), velocities vary with direction much more at high pressures. 

In conclusion, we have computed the high pressure and temperature elasticity of diamond by first principles using a novel analytical approach.  In this new powerful method, vibrational density of states for strained configurations are unnecessary \citep{Wu11}. Treatment of strain Gr\"uneisen parameters via isotropic averages reduced greatly the computational cost of the task, which, otherwise, would have been much more demanding. We have shown an outstanding improvement in the calculated elasticity of diamond as a function of temperature at ambient pressure (Figure \ref{fig_1}). These results increase the confidence in our approach to calculate high temperature elasticity (Figure \ref{fig_2}).  In general, accurate elasticity of diamond at high pressures and temperatures is a step forward in improving and opening applications of diamonds in science and technology, particularly in geophysics and planetary sciences where extreme conditions of pressure and temperature are typical. The elastic anisotropy of diamond is shown to increase considerably at high pressures. This result is significant for understanding the performance under pressure of large SC and PC diamonds specimens nowadays synthesized by CVD or as NPDs. 
\begin{acknowledgments}
We acknowledge the support of grant NSF/EAR 1047629. Computations were performed at the MSI. The authors also thank Z. Wu for helpful discussions.
\end{acknowledgments}

\bibliography{apl_diamond}

\end{document}